\documentclass[3p,a4paper,american,pdftex,twocolumn]{elsarticle}
\usepackage{amsfonts,amsmath,amssymb}
\usepackage[T1]{fontenc}
\usepackage[utf8]{inputenc}
\usepackage{graphicx}
\usepackage{microtype}
\usepackage[obeyFinal,textsize=footnotesize]{todonotes}
\usepackage{hyperref}
\usepackage[todos]{CMImacros}

\biboptions{numbers,sort&compress}
\journal{arXiv}

\newcommand{\suppinfo}{Supplementary Information\xspace}

\newcommand{\water}{\ensuremath{\HHO}\xspace}
\newcommand{\pyrrolew}{\ensuremath{\text{pyrrole}(\HHO)}\xspace}
\newcommand{\pyrrolewd}{\ensuremath{\text{pyrrole}(\HHO)_2}\xspace}
\newcommand{\pyrroled}{\ensuremath{(\text{pyrrole})_2}\xspace}

\begin{document}
\begin{frontmatter}
   \title{Spatial separation of pyrrole and pyrrole-water clusters}%
   \author[cfeldesy,uhhphys]{Melby Johny}%
   \author[cfeldesy]{Jolijn~Onvlee}%
   \author[cfeldesy,uhhphys,uhhcui]{Thomas Kierspel\fnref{fn1}}
   \author[cfeldesy,uhhphys,uhhcui]{Helen Bieker}%
   \author[cfeldesy,uhhcui]{Sebastian Trippel\corref{cor}}\cortext[cor]{Corresponding~author}%
   \ead{sebastian.trippel@cfel.de}\ead[url]{https://www.controlled-molecule-imaging.org/}%
   \author[cfeldesy,uhhphys,uhhcui]{Jochen~Küpper}%
   \fntext[fn1]{Present~address:~Department of Chemistry, University of Basel, Klingelbergstrasse
      80, Basel 4056, Switzerland}%
   \address[cfeldesy]{Center for Free-Electron Laser Science, Deutsches Elektronen-Synchrotron DESY,
      Notkestrasse 85, 22607 Hamburg, Germany}%
   \address[uhhphys]{Department of Physics, Universität Hamburg, Luruper Chaussee 149, 22761
      Hamburg, Germany}%
   \address[uhhcui]{The Hamburg Center for Ultrafast Imaging, Universität Hamburg, Luruper Chaussee
      149, 22761 Hamburg, Germany}%
   \date{\today}%
   \begin{abstract}\noindent%
      We demonstrate the spatial separation of pyrrole and \pyrrolew clusters from the other atomic
      and molecular species in a supersonically-expanded beam of pyrrole and traces of water seeded
      in high-pressure helium gas. The experimental results are quantitatively supported by
      simulations. The obtained \pyrrolew cluster beam has a purity of $\ordsim100$~\%. The
      extracted rotational temperature of pyrrole and \pyrrolew from the original supersonic
      expansion is $T_\text{rot}=0.8\pm0.2$~K, whereas the temperature of the deflected,
      pure-\pyrrolew part of the molecular beam corresponds to $T_\text{rot}\approx0.4$~K.
   \end{abstract}
   \begin{keyword}
      pyrrole, \pyrrolew cluster, Stark effect, cold molecules, electric deflection, species
      separation
   \end{keyword}
\end{frontmatter}

\section{Introduction}
\label{sec:introduciton}

Studies of solvation effects of biologically relevant aromatic molecules provide details on the
influence of the molecule's local environment on its function and on the nature of molecular
interactions, as well as insights into the functions of complex biological
systems~\cite{Colombo:Science256:655}. Prototypical model chromophores such as imidazole and pyrrole
were used to build synthetic polyamide ligands for the recognition of the Watson--Crick base pairs
in the DNA minor groove~\cite{White:Nature391:468}. Furthermore, pyrrole is a model of tryptophan's
indole chromophore, one of the strongest near-UV absorbers in proteins. Similarly, pyrrole is
responsible for the photoconversion mechanism in the phytochrome enzyme~\cite{Ulijasz:Nature463:250}
and it is a promising building block for organic dye-sensitized solar cells~\cite{Yen:JPCC112:12557}
as well as biological sensors~\cite{Tan:SABC233:599}.

A key reason for the intriguing photophysics of the above-mentioned chromophores is the excited
$^1\pi\sigma^*$ state, which is repulsive along the N-H-stretching
coordinate~\cite{Sobolewski:PCCP4:1093, Lippert:CPC5:1423, Frank:CP343:347}. The photophysics and
photochemistry of these molecules is fairly sensitive to the
environment~\cite{Sobolewski:CPL321:479}. In a bottom-up approach, spectroscopic and theoretical
investigations were performed for micro-solvated clusters to get fundamental insights into their
photophysical and photochemical properties~\cite{Lippert:CPL376:40, Sobolewski:CPL329:130,
   Lippert:CPC5:1423}. Recently performed experiments have provided evidence for ultrafast
intermolecular relaxation processes in electronically-excited microsolvated
tetrahydrofuran-water~\cite{Ren:NatPhys:79:1745} and \indolewater~\cite{Kierspel:Dissertation:2016}
clusters. This reflects one of the proposed efficient mechanisms for radiation damage processes of
biomolecules via auto-ionization caused by secondary electrons~\cite{Alizadeh:ARPC66:379}.
Time-resolved experiments, such as photoion and photoelectron spectroscopy, aiming at the
investigation of the photophysics and photochemistry of pyrrole, were performed to study the
dynamics of H elimination from the N-H site of the molecule mediated by the excitation of the
$^1\pi\sigma^*$ state~\cite{Ashfold:Science1637:1640, Roberts:FD95:115, Kirkby:CPL683:179,
   Lippert:CPC5:1423}.

For \pyrrolew clusters theoretical calculations predict that an electron is transferred across the
hydrogen bond without photodissociation of the pyrrole moiety~\cite{Frank:CP343:347,
   Sobolewski:CPL321:479}. Hence, detailed investigations of the photophysics of \pyrrolew, and
their comparison with the existing observations for \indolew~\cite{Sobolewski:CPL329:130,
   Lippert:CPC5:1423}, promises to unravel these fundamental processes in the intermolecular
interactions as well as the radiation damage of biological systems.

Advanced experiments aiming at unraveling these ultrafast dynamics often rely on pure samples of the
individual species. Such controlled samples were previously exploited in investigations of the
photophysics of indole~\cite{Kierspel:PCCP20:20205} and \indolew
clusters~\cite{Kierspel:Dissertation:2016} following site-specific soft x-ray ionization and are
even amenable to coherent-x-ray-diffraction studies~\cite{Kuepper:PRL112:083002}. These experiments
relied on the spatial separation of individual species by the electrostatic
deflector~\cite{Chang:IRPC34:557}, which was previously demonstrated for the separation of \indolew
from indole~\cite{Trippel:PRA86:033202, Chang:IRPC34:557, Trippel:RSI89:096110}.

Here, we demonstrate the spatial separation of pyrrole, a sub unit of indole and tryptophan, and the
microsolvated \pyrrolew cluster. The structure of pyrrole and \pyrrolew was studied using microwave
spectroscopy~\cite{Michael:JPC97:7451, Nygaard:JMolStruct3:491} and it was concluded that the
singly-hydrogen-bonded \pyrrolew cluster has a well defined structure, with the water attached to
the N-H site of pyrrole~\cite{Michael:JPC97:7451, Frank:CP343:347}.

\section{Experimental setup}
\label{sec:setup}
\begin{figure*}
   \includegraphics[width=1\linewidth]{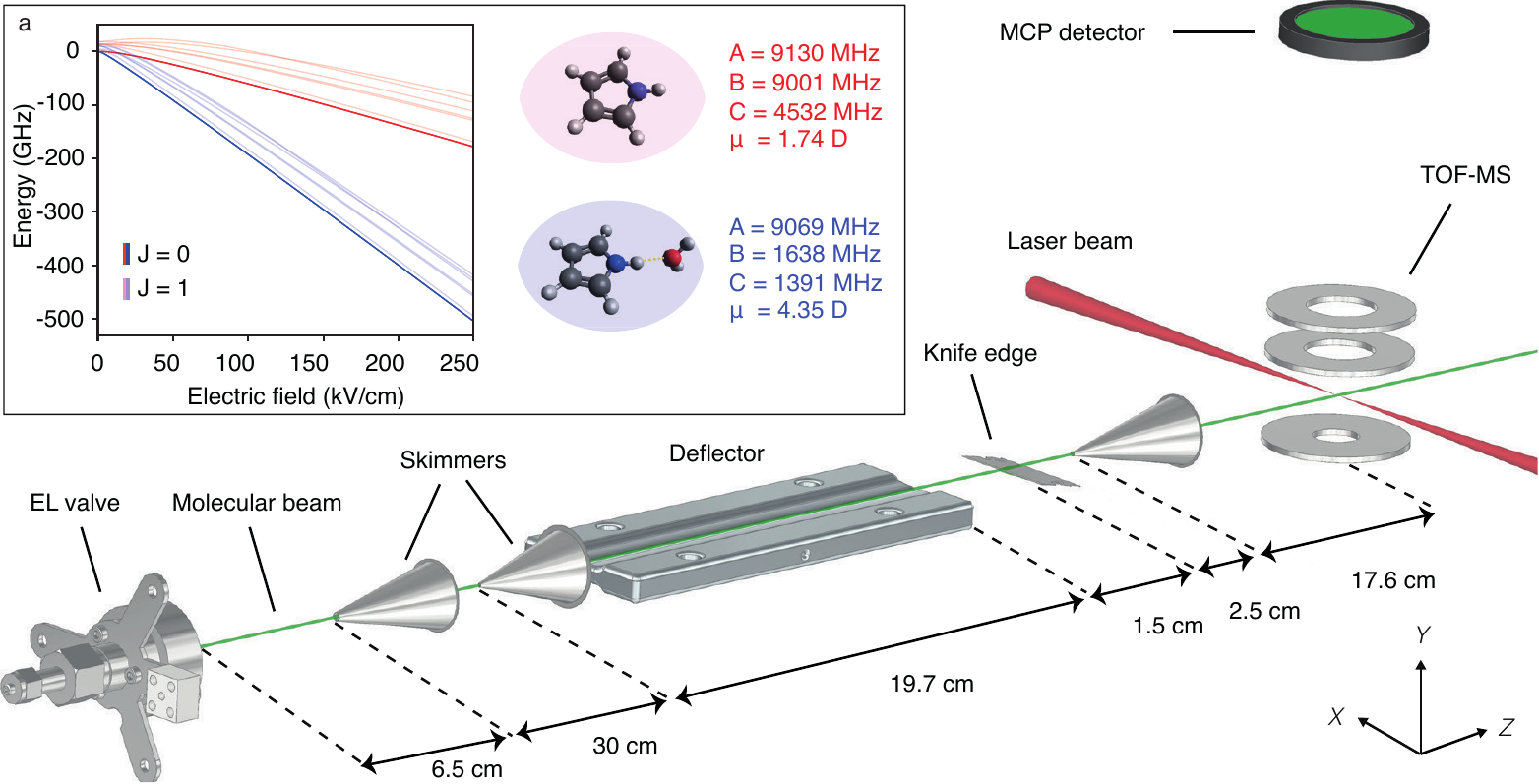}
   \caption{Schematic representation of the experimental setup and the definition of the coordinate
      system. (a) Stark energies and molecular constants of pyrrole (red) and \pyrrolew (blue).}
   \label{fig:setup}
\end{figure*}
A schematic representation of the experimental setup is shown in~\autoref{fig:setup}. An Even-Lavie
valve~\cite{Even:JCP112:8068} was used to generate a pulsed molecular beam by supersonic expansion
of a few millibar of pyrrole (Sigma Aldrich, $>98$~\%) and traces of water seeded in $\ordsim80$~bar
of helium into vacuum. The valve was operated at a repetition rate of 250~Hz and was heated to
\celsius{65}. The expanded molecular beam was then skimmed twice using conical skimmers (Beam
Dynamics, model 50.8, $\varnothing=3.0$~mm \& model 40.5, $\varnothing=1.5$~mm), which were placed
at distances of 6.5~cm and 30.2~cm downstream the nozzle, respectively. An inhomogeneous electric
field created by the so-called $b$-type deflector~\cite{Kienitz:JCP147:024304} was used to disperse
the molecular beam according to the species' effective-dipole-moment-to-mass
ratio~\cite{Chang:IRPC34:557, Filsinger:JCP131:064309, Filsinger:PRL100:133003}. The molecular beam
was cut by a vertically adjustable knife edge placed 1.5~cm downstream of the exit of the deflector,
which allowed for an improved separation of all species present in the molecular
beam~\cite{Trippel:RSI89:096110}. The experiments were conducted by placing the knife edge at a
height where it cut off the undeflected (0~kV applied on deflector electrodes) molecular beam at the
center of the vertical column density profile. The molecular beam was further skimmed by a conical
skimmer (Beam Dynamics, model 50.8 with $\varnothing=1.5$~mm) placed 4~cm downstream of the exit of
the deflector. The transverse positions of the valve, skimmers, and the deflector were adjustable
using motorized translation stages. A laser pulse with a duration of $\ordsim30$~fs, a wavelength
centered at 800~nm, focused to $\varnothing\approx50~\um$, and directed perpendicular to the
molecular beam ionized molecules in the extraction region of a time-of-flight mass-spectrometer
(TOF-MS) placed 17.6~cm downstream of the last skimmer. The peak intensity of the laser pulse was
$\ordsim2\times10^{14}$~W/cm$^2$. The ions generated due to the strong-field ionization were
detected using a micro-channel plate (MCP), operated in single-shot readout.

\section{Results and discussions}
\label{sec:discussion}
\begin{figure}[t]
   \includegraphics[width=\linewidth]{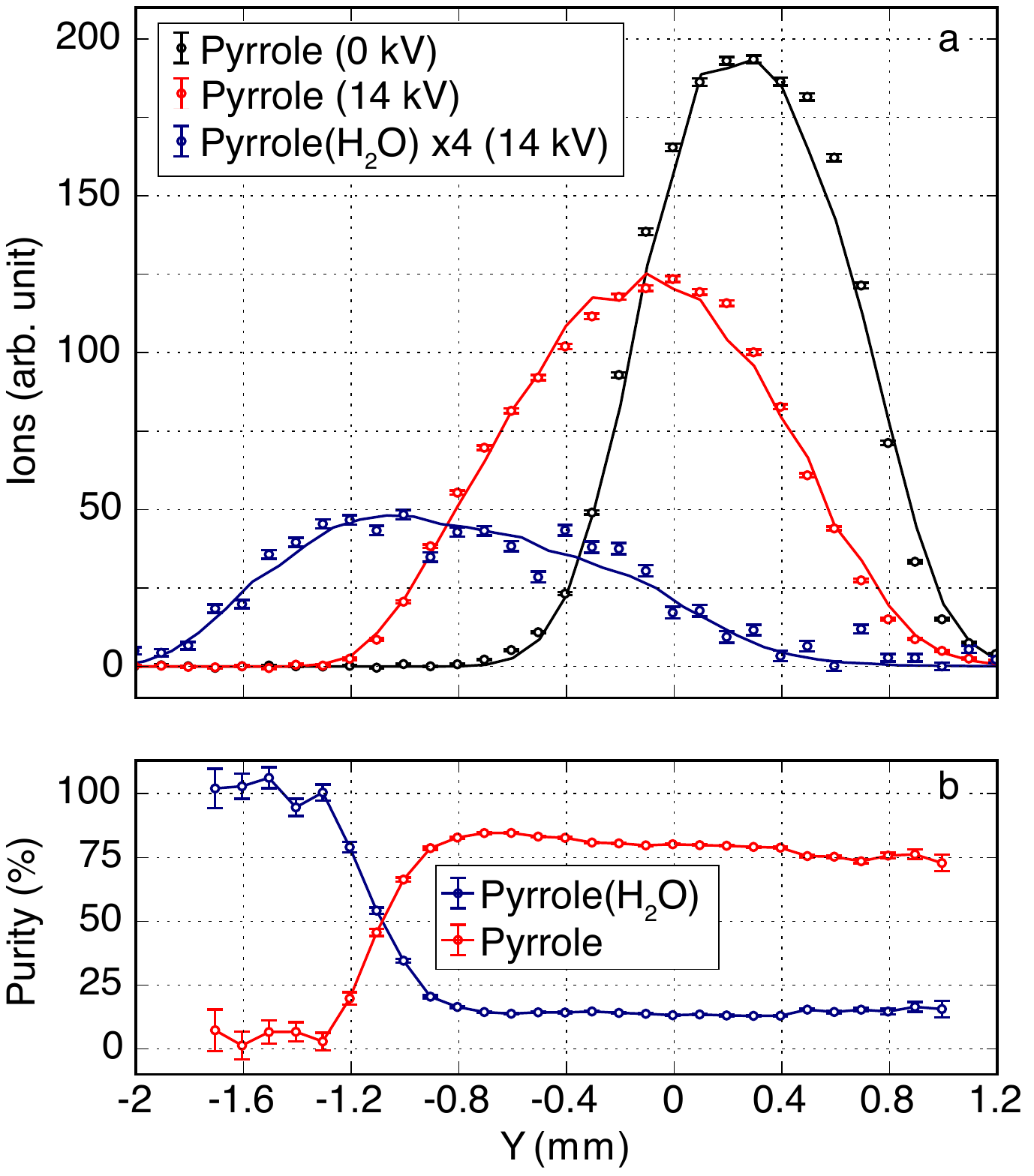}
   \caption{a) Measured vertical column density profiles for the undeflected (0 kV) pyrrole (black
      dots), deflected (14 kV) pyrrole (red dots) and \pyrrolew (blue dots). Solid lines are the
      corresponding simulated profiles. b) Purity of pyrrole (red) and \pyrrolew (blue).}
   \label{fig:deflection}
\end{figure}
\autoref[a]{fig:setup} shows the structure, rotational constants~\cite{Michael:JPC97:7451,
   Nygaard:JMolStruct3:491}, permanent dipole moment~\cite{Michael:JPC97:7451,
   Nygaard:JMolStruct3:491}, and the Stark energies of pyrrole (red curves) and \pyrrolew (blue
curves) for the $J=0,1$ rotational states. The Stark energies were calculated using
\textsc{CMIstark}~\cite{Chang:CPC185:339} within the rigid-rotor approximation, as justified by
previous experimental and theoretical work on the \indolew cluster~\cite{Trippel:PRA86:033202,
   Thesing:PRA98:053412}. Pyrrole has a $\ordsim64$~\% smaller Stark-energy shift than \pyrrolew for
the $J=0$ rotational state at an electric field strength of $E\approx200$~kV/cm. This enabled full
spatial separation of \pyrrolew from pyrrole in a cold molecular beam using the electrostatic
deflector.

The molecular beam species underwent fragmentation following the strong-field-ionization process.
Hence contributions of ionic fragments from larger clusters in specific mass gates were observed in
our spectra. A $\ordsim28$~\% probability of $\pyrrolew$ fragmenting into pyrrole$^+$ was
experimentally obtained for our specific laser pulse properties. A detailed description of the
fragmentation ratios and utilized mass gates for the specific ion signals is presented in the
\suppinfo. The fragmentation-corrected vertical column density profile for undeflected pyrrole is
shown as black dots in \autoref[a]{fig:deflection}. The undeflected profiles of \pyrrolew and other
species in the TOF-MS spectra had similar shapes and are not shown. In addition, the fragmentation
corrected deflection profiles of both, pyrrole (red dots) and \pyrrolew (blue dots), are shown for
voltages of $\pm7$~kV applied across the deflector. All quantum states are strong-field seeking at
the relevant electric field strengths experienced by the investigated molecules and clusters inside
the deflector. Therefore, all species are deflected downward, in the negative $Y$
direction~\cite{Kienitz:JCP147:024304}. The experimental deflection profile of \pyrrolew shows the
strongest deflection, down to $Y=-1.8$~mm, whereas pyrrole was only deflected down to $Y=-1.2$~mm.
Significant deflection of \water is not expected due to its small effective dipole
moment~\cite{Horke:ACIE53:11965}. Larger cluster species, \eg, \pyrrolewd and \pyrroled, are also
deflected less then pyrrole and \pyrrolew; these deflection profiles are shown and discussed in the
\suppinfo.

The solid lines in \autoref[a]{fig:deflection} are simulated profiles obtained from the results of
Monte-Carlo trajectory calculations that take into account the geometrical constraints of the
mechanical apertures in the experimental setup~\cite{Filsinger:JCP131:064309}. These simulated
deflection profiles of pyrrole and \pyrrolew matched the experimental data assuming an initial
rotational temperature of the molecular beam entering the deflector of $T_\text{rot}=0.8\pm~0.2$~K.
In the deflected part of the beam at $Y=-1.6$~mm the relative populations of the rotational states
of \pyrrolew were determined from the simulations. Although this rotational-state distribution is
non-thermal, it approximately corresponds to a thermal distribution of $0.4$~K, see \suppinfo. This
indicates that an ensemble of very cold molecules is generated using the
deflector~\cite{Holmegaard:PRL102:023001, Nielsen:PCCP13:18971} and that ultracold ensembles, even
of isolated well-defined molecular clusters, can be generated in the most deflected part of a
dispersed molecular beam.

The purities of pyrrole and \pyrrolew, as defined by the ratio of the specific signals to the sum of
signals of all other species observed is shown in \autoref[b]{fig:deflection}, see \suppinfo for
details of the analysis. This demonstrates that a molecular beam of \pyrrolew with a purity of
$\ordsim100$~\% was produced at vertical positions $-1.8~\text{mm}<Y<-1.3~\text{mm}$. The column
density profile of helium is not shown here, but is expected to be only slightly broader than the
undeflected (0~kV) profile of the molecular species, owing to its lighter mass. Furthermore, helium
is expected to be only marginally deflected by a few \um due to its small
polarizability~\cite{Kienitz:JCP147:024304}. Hence, the extracted pure molecular beam of \pyrrolew
should also be free from helium gas at $Y<-1$~mm.

\section{Conclusion}
\label{sec:conclusion}
We demonstrated the spatial separation of the \pyrrolew cluster in a molecular beam, \ie, from
pyrrole, water, larger clusters, and the seed gas. A purity of $\ordsim100$~\% of \pyrrolew was
obtained in the most deflected part of the molecular beam. Simulated deflection profiles were in
excellent agreement with the experiment. They yielded a rotational temperature of
$T_\text{rot}=0.8\pm0.2$~K in the initial molecular beam for both, pyrrole and \pyrrolew, and of
$T_\text{rot}\approx0.4$~K in the deflected pure \pyrrolew beam.

These pure beams of \pyrrolew provide a crucial ingredient for photophysics studies aiming at
time-resolved hydrogen bond formation/dissociation dynamics, \eg, in ultrafast laser pump and x-ray
probe experiments. Further interesting aspects will be the control of the orientation in the
laboratory frame by laser aligning or mixed-field orienting \pyrrolew
clusters~\cite{Holmegaard:PRL102:023001, Trippel:JCP148:101103, Thesing:PRA98:053412}. The separated
pure species are also ideally suited for experiments to image the structure and dynamics of the
\pyrrolew complex in the molecular frame, \eg, through molecular-frame photoelectron angular
distributions (MFPADs), gas-phase x-ray diffraction, or laser-induced electron diffraction (LIED)
experiments~\cite{Holmegaard:NatPhys6:428, Blaga:Nature483:194, Kuepper:PRL112:083002,
   Yang:PRL117:153002, Trabattoni:cutoff:inprep}.

\section*{Acknowledgments}
This work has been supported by the European Union's Horizon 2020 research and innovation program
under the Marie Skłodowska-Curie Grant Agreement 641789 ``Molecular Electron Dynamics investigated
by Intense Fields and Attosecond Pulses'' (MEDEA), by the excellence cluster ``The Hamburg Center
for Ultrafast Imaging -- Structure, Dynamics and Control of Matter at the Atomic Scale'' of the
Deutsche Forschungsgemeinschaft (CUI, DFG-EXC1074), by the European Research Council under the
European Union's Seventh Framework Program (FP7/2007-2013) through the Consolidator Grant COMOTION
(ERC-Küpper-614507), and by the Helmholtz Association through the Virtual Institute 419 ``Dynamic
Pathways in Multidimensional Landscapes'' and the ``Initiative and Networking Fund''. J.O.\
gratefully acknowledges a fellowship by the Alexander von Humboldt Foundation.

\section*{References}
\bibliography{string,cmi}
\end{document}


\begin{frontmatter}
   \title{Supplementary information: Spatial separation of pyrrole and pyrrole-water clusters}%
   \author[cfeldesy,uhhphys]{Melby Johny}%
   \author[cfeldesy]{Jolijn~Onvlee}%
   \author[cfeldesy,uhhphys,uhhcui]{Thomas Kierspel\fnref{fn1}}%
   \author[cfeldesy,uhhphys,uhhcui]{Helen Bieker}%
   \author[cfeldesy,uhhcui]{Sebastian Trippel\corref{cor}}\cortext[cor]{Corresponding~author}%
   \ead{sebastian.trippel@cfel.de}\ead[url]{https://www.controlled-molecule-imaging.org/}%
   \author[cfeldesy,uhhphys,uhhcui]{Jochen~Küpper}%
   \fntext[fn1]{Present~address:~Department of Chemistry, University of Basel, Klingelbergstrasse 80, Basel 4056, Switzerland}%
   \address[cfeldesy]{Center for Free-Electron Laser Science, Deutsches Elektronen-Synchrotron DESY,
      Notkestrasse 85, 22607 Hamburg, Germany}%
   \address[uhhcui]{The Hamburg Center for Ultrafast Imaging, Universität Hamburg, Luruper Chaussee
      149, 22761 Hamburg, Germany}%
   \address[uhhphys]{Department of Physics, Universität Hamburg, Luruper Chaussee 149, 22761
      Hamburg, Germany}%
   \date{\today}%
\end{frontmatter}

\begin{figure}[b]
   \includegraphics[width=\linewidth]{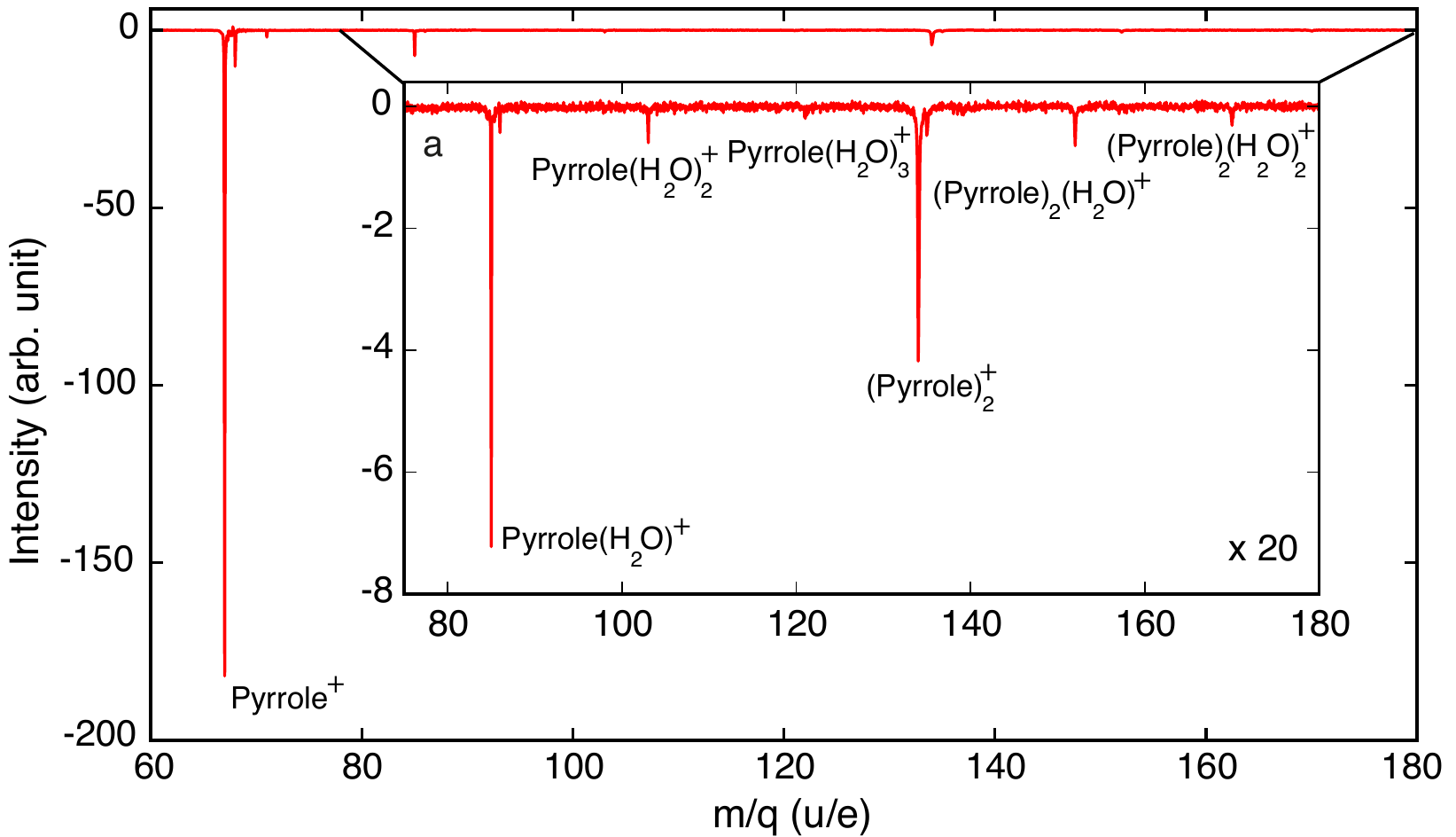}
   \caption{The mass spectra of pyrrole monomer, and larger clusters of pyrrole and water. a) Zoom
      in of the mass spectrum for clusters.}
   \label{fig:TOF}
\end{figure}

The mass spectrum of the molecular beam after strong field ionization, recorded in the center of the
undeflected (0~kV) molecular beam profile, is shown in~\autoref{fig:TOF}.
In the spectrum, mass peaks that correspond to pyrrole$^+$, $\pyrrolew^+$, $\pyrrolewd^+$,
$\pyrroled^+$, $\pyrroledw^+$, $\pyrroledwd^+$, $\pyrrolewt^+$, and larger clusters were
observed. Column-densities were obtained by integrating the mass spectrum within gates $\pm0.5$~u/e
around the nominal masses of pyrrole$^+$, $\pyrrolew^+$, $\pyrroled^+$, and $\pyrrolewd^+$; the
resulting vertical beam profiles are shown in \autoref[a]{fig:deflection}.
\begin{figure}
   \includegraphics[width=\linewidth]{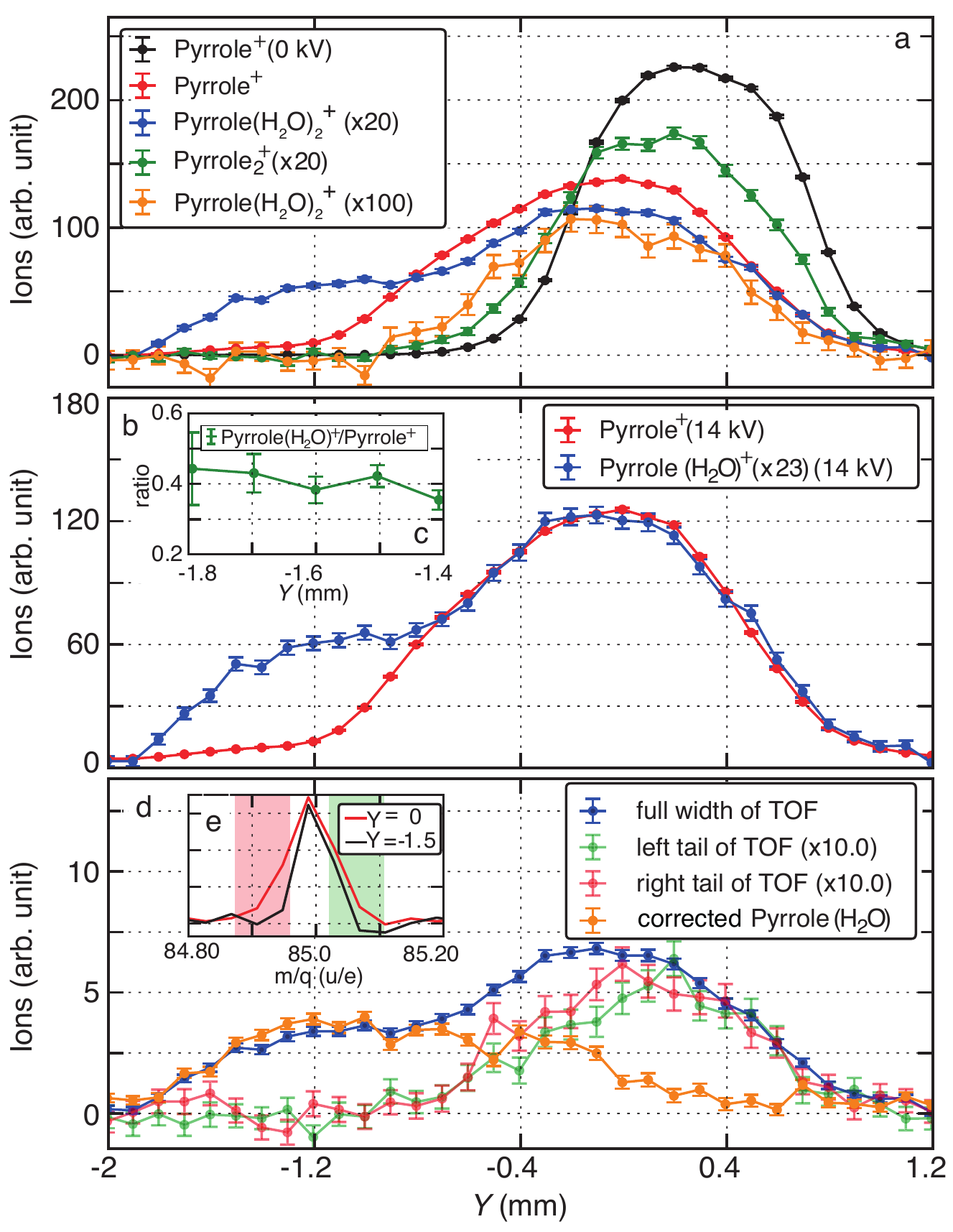}
   \caption{a) Vertical column density profiles recorded for various ion channels. b) Vertical
      column density profiles of deflected (14 kV) pyrrole$^+$ and $\pyrrolew^+$. c) Ratio of
      $\pyrrolew^+$/pyrrole$^+$ at positions $-1.8~\text{mm}<Y<-1.4$~mm. d) Molecular beam profile
      of \pyrrolew with gates covering the full-width of the TOF-MS peak (blue, $m/q=85\pm~0.5$~u/e)
      and curves corresponding to the vertical profile of \pyrrolew with gates set to the wings of
      TOF-MS peak (red and green), \ie, the red ($m/q= 84.88$--$84.96$~u/e) and green
      ($m/q= 85.02$--$85.10$~u/e) shaded regions shown in the inset, e). The orange curve
      corresponds to the fragmentation corrected column density profile of \pyrrolew. e) TOF-MS of
      deflected (14~kV) $\pyrrolew^+$ at vertical position $Y=0$ (red) and of pure \pyrrolew beam at
      $Y=-1.5$~mm (black, $m/q=85\pm~0.5$~u/e).}
   \label{fig:deflection}
\end{figure}
The plotted profiles are for undeflected pyrrole$^+$ (0 kV) and deflected (14 kV) molecular-beam
species without any fragmentation correction. The profiles of all species were scaled to the
undeflected profile of pyrrole$^+$ for an improved visibility, see the legend of
\autoref[a]{fig:deflection} for the scaling factors used in the plot. The amount of deflection was
in the following order: $\pyrrolew^+>\text{pyrrole}^+>\pyrrolewd^+>\pyrroled^+$. Water is expected
to deflect less than \pyrroled because of its small effective dipole moment at the electric field
strengths used in the experiment. The fragmentation of $\pyrrolew^+$ into the pyrrole$^+$ ion is
evident from the matching shape of the deflection profile of pyrrole$^+$ with that of $\pyrrolew^+$
at vertical positions \mbox{$-1.8~\text{mm}<Y<-1.4~\text{mm}$}. This manifests itself in a constant
ratio \mbox{$R=S(\pyrrolew^+)/S(\text{pyrrole}^+)$} of the specific integrated signals $S(j)$ of the
species $j$ in that region as shown in \autoref[c]{fig:deflection}. We have corrected for the
contribution of the fragmentation of $\pyrrolew^+$ into the pyrrole$^+$ channel using the mean value
of the ratio $R$ between vertical positions $-1.8~\text{mm}<Y<-1.4~\text{mm}$,
\mbox{$R=S(\pyrrolew^+)/S(\text{pyrrole}^+)=0.38$}.

The deflection profile of \pyrrolew was also corrected for the fragmentation from larger clusters.
For this the width of the TOF-MS peak of pure $\pyrrolew^+$ in the most deflected part of the beam
($Y=-1.5$~mm) was compared with the width of the $\pyrrolew^+$ peak in the center ($Y=0$~mm) of the
molecular beam profile where we have contribution from higher clusters, \autoref[e]{fig:deflection}.
The TOF-MS of $\pyrrolew^+$ with contribution from larger clusters (red curve) was broader than the
one of the deflected pure $\pyrrolew^+$ beam (black curve). This is attributed to the kinetic-energy
release upon cluster fragmentation; analyzing this allowed us to correct for these contributions.
The vertical column density profiles in \autoref[d]{fig:deflection} obtained by setting gates to the
wings of the TOF-MS peak (green line, $m/q=85.02$--$85.10$~u/e and red line,
\mbox{$m/q=84.88$--$84.96$~u/e}) were scaled to the least deflected part of the $\pyrrolew^+$ signal
($m/q=85\pm0.5$~u/e) at $0.2~\text{mm}<Y<1$~mm, with a scaling factor of $10$. The fragmentation
corrected $\pyrrolew^+$ profile (orange line) was obtained by subtracting these
larger-cluster-fragment corrections (green and red lines) from the profile obtained by setting gates
to the full width of the TOF-MS peak (blue line, $m/q=85\pm0.5$~u/e). The step at vertical positions
ranging from $Y=-0.3$ to $-0.8$~mm (blue line) and the shape of the profiles for larger clusters
(red and green lines) matched the shape of the deflected $\pyrrolewd^+$ in
\autoref[a]{fig:deflection}. Hence, the uncorrected $\pyrrolew^+$ profile had significant
contributions from fragments of larger, weakly deflected, $\pyrrolewd^+$ cluster into the
$\pyrrolew^+$ channel.

The purity $P_i$ of a species $i$ is defined by \mbox{$P_i=S(i^+)/\sum_j{S(j^+)}$}. For our
evaluation of the purity, integrated signals from pyrrole$^+$, $\pyrrolew^+$, $\pyrrolewd^+$,
$\pyrroled^+$, $\pyrroledw^+$, and $\pyrroledwd^+$ were taken into account; larger clusters were
neglected.

\bigskip

\begin{figure}
   \includegraphics[width=1\linewidth]{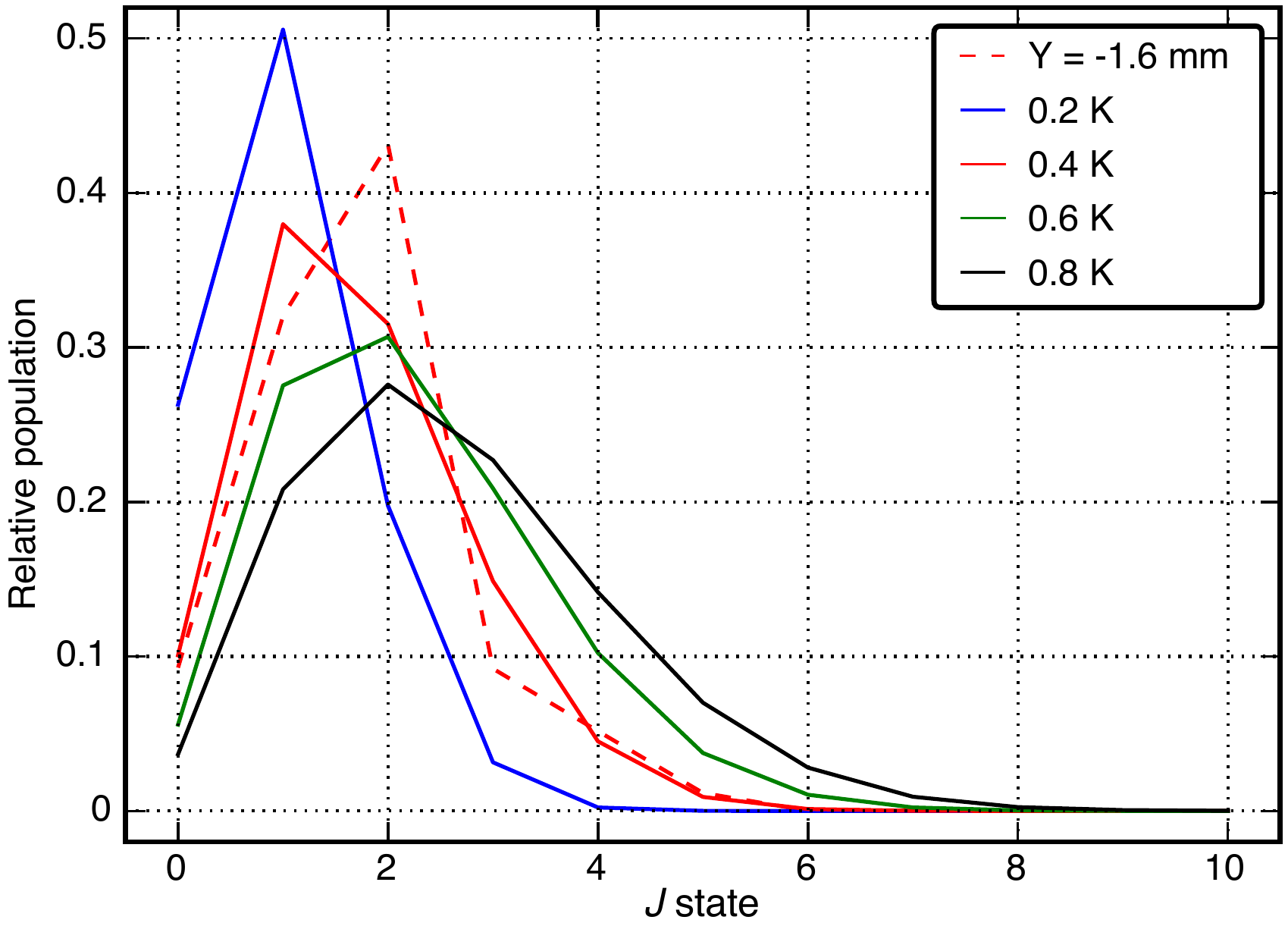}
   \caption{The relative population of the rotational states in the molecular beam at $Y=-1.6$~mm
      (dashed line) and the thermal distributions at various temperatures (solid lines).}
   \label{fig:population}
\end{figure}

The relative populations of rotational states of \pyrrolew in the molecular beam at
\mbox{$Y=-1.6$~mm} were determined from the simulations, see \autoref{fig:population}; this
distribution is non-thermal due to the nature of the deflection process. The rotational state
distribution for a range of temperatures are also shown in the plot. The relative populations in the
deflected part of the beam match fairly well with a thermal distribution of
$T_\text{rot}=0.4$~K. This confirms that ultracold molecular beams, even of isolated well-defined
molecular clusters, can be generated in the most deflected part of a dispersed molecular beam.
